# Self-Lensing By Binaries

**Andrew Gould**[*]

Dept of Astronomy, Ohio State University, Columbus, OH 43210

e-mail gould@payne.mps.ohio-state.edu

**Abstract**

Lensing of one member of a binary by its companion is studied for several classes of binaries. For binaries in which at least one member is an ordinary (non-compact) star, the optical depths to lensing is extremely low, $\tau \lesssim v_{\rm esc}^4/c^4 \sim 10^{-11}$. Here $v_{\rm esc}$ is the escape velocity from the component with the weaker potential. Binaries composed of two compact objects can have higher optical depths. About 0.2% of millisecond pulsar binaries will be self-lensing. Such objects can be used to probe the pulsar-emission mechanism.

Subject Headings: binaries – gravitational lensing – pulsars

---

## 1. Introduction

Gravitational microlensing occurs when a mass (the lens) comes close to the observer's line of sight to a source. Three ongoing experiments are searching for gravitational microlensing toward the Large Magellanic Cloud (LMC) (Alcock et al. 1993; Aubourg et al. 1993) and toward the galactic bulge (Udalski et al. 1993; Alcock et al. 1994). These experiments find that the optical depth is $\tau \sim \mathcal{O}(10^{-7})$ toward the LMC and $\tau \sim \mathcal{O}(10^{-6})$ toward the bulge. The optical depth is defined as the fraction of the sky covered by the "Einstein rings" of the lenses. The Einstein radius is

$$r_e = \left(\frac{4GmD_{\mathrm{OL}}D_{\mathrm{LS}}}{c^2 D_{\mathrm{OS}}}\right)^{1/2}, \qquad (1.1)$$

where $m$ is the mass of the lens and $D_{\mathrm{OL}}$, $D_{\mathrm{LS}}$, and $D_{\mathrm{OS}}$ are the distances between the observer, lens, and source.

A question that naturally arises in relation to these experiments is: what fraction of events are due to lensing of the source star by a binary companion? Here I show that the optical depth to self-lensing by binaries is negligibly small, $\tau \lesssim v_{\mathrm{esc}}^4/c^4 \sim 10^{-11}$, where $v_{\mathrm{esc}}$ is the escape velocity from the binary member with the smaller potential. However, the fact that the limit to the optical depth scales as a high power of the escape velocity leads me to investigate self-lensing by binaries of compact objects.

I find that millisecond pulsar binaries present the best prospect for detecting self-lensing. The fraction of such binaries that give rise to lensing events is $\sim 0.2\%$. It is relatively easy to determine whether a particular pulsar is a lensing candidate and to predict the phase of the orbital period when the lensing event should take place. If observed, such systems could be used to probe the geometry of the pulsar mechanism.

# 2. Self-Lensing By Binaries

## 2.1. POINT SOURCES

Consider a binary composed of two masses, $m_1$ and $m_2$, and orbital separation $r(t)$. Initially, I assume that the components are point masses and point sources. If the system is viewed from a distance $D_{\rm OS} \gg r(t)$ with $m_1$ approximately in front of $m_2$, then from equation (1.1) the Einstein radius for lensing $m_2$ is

$$r_{e,1}(t) = \frac{[4Gm_1 r(t)]^{1/2}}{c}. \tag{2.1}$$

Consider now an ensemble of such systems seen at random viewing angles, but all with the same orbital separation as a function of time. The optical depth to lensing as a function of time $\tau(t)$ can be evaluated

$$\tau_1(t) = \frac{\pi [r_{e,1}(t)]^2}{4\pi [r(t)]^2} = \frac{Gm_1}{c^2 r(t)}. \tag{2.2}$$

Hence the optical depth $\tau_1$ time-averaged over the orbit is

$$\tau_1 = \langle \tau_1(t) \rangle = \frac{Gm_1}{ac^2} = \frac{m_1}{M} \frac{\langle v^2 \rangle}{c^2}, \tag{2.3}$$

where $M \equiv m_1 + m_2$ is the total mass of the system, $\langle v^2 \rangle$ is the mean square relative velocity, and $a$ is the semi-major axis. A similar result applies for $\tau_2$, the optical depth for lensing of $m_1$ by $m_2$. The total optical depth, $\tau = \tau_1 + \tau_2$ is then

$$\tau = \frac{GM}{ac^2} = \frac{\langle v^2 \rangle}{c^2}. \tag{2.4}$$

## 2.2. FINITE SOURCES

I now consider that the components have finite radii, $R_1$ and $R_2$. For simplicity, I assume that the orbit is circular with relative speed $v$. I again focus on lensing of $m_2$ by $m_1$. If the Einstein radius is smaller than the source size, $r_{e,1} < R_2$, then the maximum magnification due to lensing will be suppressed to a level $A_{\max} \sim 1 + 2r_{e,1}^2/R_2^2$ (Gould 1992). If $r_{e,1} < R_1$ then the source will be occulted. That is the system would be an eclipsing rather than a lensing binary. Hence lensing can be significant only if $\max\{r_{e,1}, r_{e,2}\} > R$, where $R = \max\{R_1, R_2\}$. Using equations (2.2) and (2.3), this implies

$$\tau \lesssim \left(\frac{2Gm}{Rc^2}\right)^2, \qquad (2.5)$$

where $m \equiv \max\{m_1, m_2\}$. For systems with roughly equal masses, equation (2.5) may be rewritten

$$\tau \lesssim \frac{v_{\rm esc}^4}{c^4}, \qquad (2.6)$$

where $v_{\rm esc}$ is the escape velocity from the surface of the component with the weaker potential. Equation (2.6) shows that binaries in which either component is an ordinary star have optical depths $\tau \lesssim 10^{-11}$. Lensing by such binaries is completely negligible.

## 3. White Dwarfs

For a binary composed of two white dwarfs with $R \sim 3\,R_\oplus$ and $m \sim 0.6\,M_\odot$, the maximum optical depth would be $\tau \sim 10^{-8}$, the value when $a \sim 1\,{\rm AU}$. This optical depth is small. However, it should be noted that a fraction $f \sim R/a \sim 10^{-4}$ of such systems are lensing binaries. The lensing events have time scales $\omega^{-1} \equiv r_e/v \sim 10\,{\rm min}$, compared to an orbital period $P_{\rm orb} \sim 1\,{\rm yr}$. Thus, the lensing event lasts a fraction $(\omega P_{\rm orb})^{-1} \sim 2 \times 10^{-5}$ of the orbit.

From the above calculation, $\gtrsim 10^5$ observations would have to be made of each of $\gtrsim 10^4$ white-dwarf binaries in order to find a single lensing binary.

# 4. Neutron Stars

For both neutron-star and black-hole binaries, a naive application of equation (2.6) leads to very high estimates of possible optical depths. One quickly finds however that the resulting systems are unstable to collapse brought on by emission of gravitational waves. For many applications, it is natural to assume that the system is stable for some time $T$. For example, if millisecond pulsars are being lensed, $T$ would be $\sim 1\,\mathrm{Gyr}$, the estimated lifetime of such pulsars.

## 4.1. Limit From Gravitational Radiation

The mean power output of a system by gravitational radiation is

$$\left\langle \frac{dE}{dt} \right\rangle = -\frac{1}{5} \sum_{a,b} \left\langle \left( \frac{d^3 I_{ab}}{dt^3} \right)^2 \right\rangle, \tag{4.1}$$

where $I_{ab}$ is the moment of inertia tensor and where for simplicity I use $G = c = 1$ (Misner, Thorne, & Wheeler 1973). Assuming circular orbits, this yields $dE/dt = -(32/5) m_1^2 m_2^2 M a^{-5}$. Since $E = -m_1 m_2 / 2a$, I find that the minimum radius $a_*$ for which the system can survive a lifetime $T$ is related to the mass of the system $M$ by

$$\frac{GM}{a_* c^2} = \left( \frac{5}{256} \frac{M^3}{m_1 m_2 T} \right)^{1/4} = 7.4 \times 10^{-7} \left( \frac{m_1}{M} \frac{m_2}{M} \frac{T}{T_0} \right)^{-1/4} \left( \frac{M}{M_\odot} \right)^{1/4}, \tag{4.2}$$

where I define $cT_0 \equiv 10^{28}\,\mathrm{cm}$.

## 4.2. Optical Depth

For neutron-star binaries ($m_1 = m_2 = 1.4 M_\odot$), equations (2.4) and (4.2) imply an optical depth

$$\tau = \frac{GM}{a_* c^2} = 2 \times 10^{-6} \left( \frac{T}{0.1 T_0} \right)^{-1/4}, \tag{4.3}$$

corresponding to a minimum radius $a_* = 2 \times 10^{11}\,\mathrm{cm}$. At this minimum radius, lensing events have a characteristic time $\omega^{-1} = 10\,\mathrm{s}$ and $P_{\mathrm{orb}} \sim 8\,\mathrm{hr}$, similar to the periods of the two known systems (Taylor et al. 1992). A fraction $f \sim r_e/a_* \sim 0.2\%$ of such systems are lensing binaries. Thus $\gtrsim 2000$ observations of each of $\gtrsim 500$ systems would have to be made to find one lensing binary.

### 4.3. Millisecond Pulsars

However, the neutron-star binaries of practical interest are millisecond pulsars. Indeed, no other neutron-star binaries have ever been observed and it is difficult to imagine how they would be. It is not necessary to monitor millisecond pulsars repeatedly to search for a lensing event. Rather, the orbital solution tells one when the two components are most closely lined up along the line of sight. It is then possible to search for lensing events during the precise $\sim 10\,\mathrm{s}$ interval when they might occur. In fact, the orbital solution gives an approximate angle of inclination of the orbit, so it is possible to decide in advance which systems might be lensing.

Unfortunately, to date only two such systems are known (Taylor et al. 1992) whereas several hundred would be needed to find even one lensing system. Nevertheless, it is possible that search techniques will improve and that many more will be discovered in the future. If so, lensing binaries could be found among them.

Lensing pulsar binaries could be used to probe the structure of the emitting region of pulsars, in particular to check whether emission originates from the light cylinder or from a much smaller locus. For pulsars of period $P$, the light cylinder has a radius $Pc/2\pi \sim 10^8\,\mathrm{cm}(P/20ms)$. If the lensing magnification curve followed the classic form for a point source, this would show that the source is small on scales of the Einstein ring, $r_e \sim 3 \times 10^8\,(a/10^{11}\mathrm{cm})^{1/2}$ cm.

In addition, the relative time delay between the two lensed images of the pulsar would be

$$\Delta t \sim 8\frac{Gm}{c^3}\frac{s}{r_e} \sim 5.6 \times 10^{-2}\frac{s}{r_e}\,\mathrm{ms}, \qquad (4.4)$$

where $s$ is the projected separation of the neutron stars. This would lead to a small, but perhaps measurable distortion of the pulse during the lensing event.

## 4.4. Black-Hole Binaries

The same formalism used to analyze neutron-star binaries can also be applied to black-hole binaries. For example, if the primary black hole were $m_1 \sim 10^9 \, M_\odot$ and the secondary were smaller by a factor $\eta$, the optical depth to lensing the larger black hole would be $\tau_2 \sim 1.3 \times 10^{-4} \eta^{3/4}$ at a minimum radius $a_* \sim 10^{18} \eta^{-1/4}$ cm, and $P_{\rm orb} \sim 600 \, \eta^{-3/8}$ yr. The events would last $\omega^{-1} \sim 2\eta^{-1/4}$ yr. The low optical depths of such systems taken together with the long intervals required between observations of the same system in themselves make the problem of detecting self-lensing by black-hole binaries a formidable one. Moreover, it is not at all clear what effects of such a system might be both observable and subject to lensing. Hence, at the present time self-lensing by black-hole binaries is a speculative prospect at best.

# REFERENCES

Alcock, C., et al. 1993, Nature, 365, 621

Alcock, C., et al. 1994, ApJ, submitted

Aubourg, E., et al. 1993, Nature, 365, 623

Gould, A. 1992, ApJ, 392, 442

Misner, C. W., Thorne, K. S., & Wheeler, J. A. 1973, Gravitation, (San Francisco: Freeman)

Udalski, A., Szymański, J., Kaluzny, J., Kubiak, M., Krzemiński, W., Mateo, M., Preston, G. W., & Paczyński, B. 1993, Acta Astronomica, 43, 289

Taylor, J. H., Wolszczan, A., Damour, T., & Weisberg, J. M. 1992, Nature, 355, 132